\documentclass[aps,prc,superscriptaddress,twoside,twocolumn,nofootinbib,showpacs]{revtex4-1}
\usepackage{amsmath,amssymb}
\usepackage{graphicx,bm}
\usepackage{slashed}
\usepackage{dcolumn}
\usepackage{epstopdf}
\usepackage{ulem}
\usepackage[usenames,dvipsnames]{color}
\usepackage{subfigure}
\usepackage{tensor}

\begin{document}

\title{A study of the almost sequential mechanism of true ternary fission}

\author{R. B. Tashkhodjaev}
\email{tashkhodjaev@inp.uz}
\affiliation{Institute of Nuclear Physics, Uzbek Academy of Science,
100214 Tashkent, Uzbekistan}
\affiliation{Inha University in Tashkent, 100170, Tashkent, Uzbekistan}

\author{A. I. Muminov}
\affiliation{Institute of Nuclear Physics, Uzbek Academy of Science,
100214 Tashkent, Uzbekistan}

\author{A. K. Nasirov}
\email{nasirov@jinr.ru}
\affiliation{Institute of Nuclear Physics, Uzbek Academy of Science,
100214 Tashkent, Uzbekistan}
\affiliation{Joint Institute for Nuclear Research, Joliot-Curie 6, 141980 Dubna, Russia}
\affiliation{Department of Physics, Kyungpook National University, Daegu 702-701, Korea}

\author{W. von Oertzen}
\affiliation{Helmholtz-Zentrum Berlin, Glienickerstr. 100, 14109 Berlin, Germany}
\affiliation{Fachbereich Physik, Freie Universit{\"a}t, Berlin, Germany}

\author{Yongseok Oh}
\email{yohphy@knu.ac.kr}
\affiliation{Department of Physics, Kyungpook National University, Daegu 702-701, Korea}
\affiliation{Asia Pacific Center for Theoretical Physics, Pohang, Gyeongbuk 790-784, Korea}

\begin{abstract}
We consider the collinear ternary fission which is a sequential ternary decay with
a very short time between the ruptures of two necks connecting the middle cluster of
the ternary nuclear system and outer fragments.
In particular, we consider the case where the Coulomb field of the first massive fragment
separated during the first step of the fission produces a lower pre-scission barrier in the second
step of the residual part of the ternary system.
In this case, we obtain a probability of about $10^{-3}$ for the yield of massive clusters such as
\nuclide[70]{Ni}, \nuclide[80-82]{Ge}, \nuclide[86]{Se}, and \nuclide[94]{Kr} in the ternary fission of
\nuclide[252]{Cf}.
These products appear together with the clusters having mass numbers of $A = 132$--$140$.
The results show that the yield of a heavy cluster such as \nuclide[68-70]{Ni}
would be followed by a product of $A = 138$--$148$ with a large probability
as observed in the experimental data obtained with the FOBOS spectrometer at the Joint Institute
for Nuclear Research. The third product is not observed.
The landscape of the potential energy surface shows that the configuration of the
$\nuclide{Ni} + \nuclide{Ca} + \nuclide{Sn}$ decay channel is lower about 12~MeV than
that of the $\nuclide{Ca} + \nuclide{Ni} + \nuclide{Sn}$ channel.
This leads to the fact, that the yield of \nuclide{Ni} and \nuclide{Sn} is large.
The analysis on the dependence of the velocity of the middle fragment on mass numbers
of the outer products leads to the conclusion that, in the collinear tripartition channel of
\nuclide[252]{Cf}, the middle cluster has a very small velocity, which does not allow it to be
found in experiments.
\end{abstract}

\keywords{fission, potential energy surface, ternary fission}

\pacs{
21.60.Gx, 
25.85.Ca  
}

\maketitle


\section{Introduction}

True ternary fission of heavy nuclei, which has been discovered in the experiments
recently, occurs with much smaller probability ($\sim 10^{-3}$) compared to the binary fission
~\cite{PKVA10,PKKA10,PKAA11,PKVA12}.
These experimental studies of the decays in
$^{252}$Cf(sf,fff) \cite{PKVA10} and $^{235}$U(n$_{\rm th}$,fff) \cite{PKVA12} reactions
 with two fission fragment coincidences
with two FOBOS-detectors \cite{PKVA10} placed at $180^{\circ}$,
using the missing mass approach, have established
the phenomenon of collinear cluster tripartition (CCT). The third product is not observed.
Only recently its dynamical properties could be investigated and it has been
 concluded that it proceeds collinearly.
More details can be found, for example, in Refs.~\cite{DG74,Royer95,MB11}, and it is expected
that the investigation of true ternary fission will allow us to extend our knowledge about
fusion-fission processes.
In the present work, we consider one of the dominant modes of the CCT.

Because of small probability, there are only a few experimental measurements of CCT as
given in Refs.~\cite{PKVA10,PKKA10,PKAA11,PKVA12}.
Moreover, theoretical studies about ternary fission are very limited and
some early works on this topic can be found, for example, in
Refs.~\cite{MG76,DG74,Royer95,PGG04}.
In these early works it was found that the ternary fission process in heavy nuclei occurs
preferably in collinear geometry, which was confirmed by the recent theoretical studies in
Refs.~\cite{MB11,VVB12,TNS11}.
In Ref.~\cite{MB11}, the authors studied the difference between equatorial configurations and collinear
configurations in ternary fission by calculating the potential energies for geometries of three
fragments touching each other, i.e., a tri-nuclear system (TNS), where only the mass number of the
central nucleus changes.
On the other hand, the kinetic energies of the CCT products were evaluated in Ref.~\cite{VVB12}
to find that in most cases the velocity of the central fragment can be very small.
This may be responsible for missing the third product in the experiments of
Refs.~\cite{PKVA10,PKKA10,PKAA11,PKVA12}.
In Ref.~\cite{TNS11} the CCT process has been considered as two sequential binary fissions.
Namely, in the first stage the excited compound nucleus decays into two fragments in an asymmetric
channel, then the heavier fragment decays further into two fragments.
As a result, three fragments are obtained with comparable masses.
In Ref.~\cite{TNS11} only the yield of ternary fission fragments with comparable masses has been
considered as it is similar to the case observed in the experiments performed by the FOBOS
group~\cite{PKKA10}.
The theoretical results of the yield of \nuclide[80]{Ge} and \nuclide[84]{Se} isotopes as the first step
for the CCT products and the products \nuclide[70]{Ni}, \nuclide[74,76]{Zn}, and \nuclide[82]{Ge}
in the second step in a sequential fission process are in good agreement with some of the
corresponding experimental data on the mass distributions of the \nuclide[252]{Cf} decay.
This observation leads to the conclusion that these events can be associated with the sequential two
step mechanism of CCT.
However, the yields of \nuclide[70]{Ni},  \nuclide[82]{Ge}, and \nuclide[84]{Se}
 in coincidence with ternary fission
masses $A = 130$--$150$ observed in Ref.~\cite{PKVA12} with a large probability were not fully
explained in Ref.~\cite{TNS11}.

\begin{figure}
\begin{center}
\includegraphics[width=0.3\textwidth]{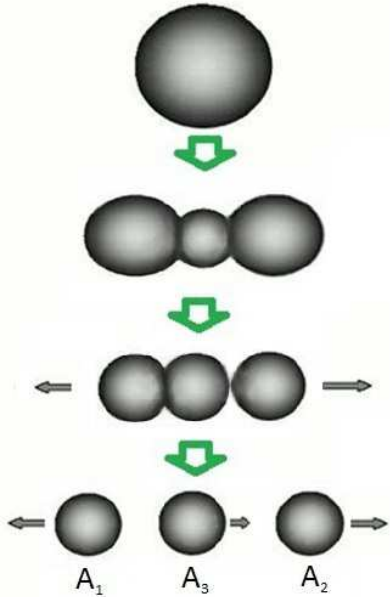}
\end{center}
\vspace*{-0.5 cm}
\caption{(Color online) The CCT fission mechanism of a heavy nucleus in
a sequential decay~\cite{TNS11}.
$A_1$, $A_2$, and $A_3$ are mass numbers of the fragments formed in the tri-nuclear system.}
\label{graph1}
\end{figure}

Therefore, in the present paper we consider the mechanism of a sequential ternary fission with a
very short time between the ruptures of the two necks connecting the middle cluster of the collinear
TNS with its outer fragments.
This mechanism is the almost-simultaneous ternary fission as illustrated in Fig.~\ref{graph1}.
The main goal of the present work is to pursue theoretical analysis of the ternary fission channels
leading to the formation of the products of mass number $A = 132$--$140$.

\begin{figure}
\begin{center}
\includegraphics[width=0.65\textwidth]{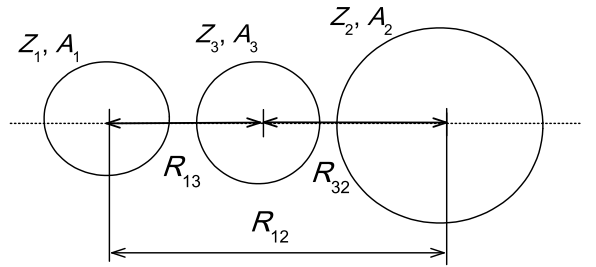}
\vspace*{-4.0cm}
\end{center}
\caption{The variables of the tri-nuclear system used in the analysis of the interaction
energy between its fragments.
Here, $Z_i$ is the charge number of fragment $i$ ($i=1,2,3$) and $R_{ij}$ is the distance between
the mass centers of fragments $i$ and $j$.
}
\label{ThreeSph}
\end{figure}

The collinear configuration of the tri-nuclear system undergoing fission is defined as follows.
First, the three fragments are situated on \textit{one line} and the border nuclei are numbered as
``1'' and ``2'', while the middle nucleus is labeled as ``3'' as shown in Fig.~\ref{ThreeSph}.
Consequently, there is no nuclear interaction between the outer fragments ``1'' and ``2''.
However, the Coulomb interaction between them is taken into account because of its long-range
property.
In fact, it was found to have a nontrivial role in the decay of TNS.
The pre-scission barrier between fragments ``1'' and ``3'' decreases due to the Coulomb field
of the fragment ``2''.
For example, in the case of the sequential ternary fission of \nuclide[236]{U}, when \nuclide[132]{Sn}
forms as the fragment ``2'', the pre-scission barrier between the fragments ``2''
 and ``3'' is smaller than
the one between the fragments ``1'' and ``3''~\cite{TNS11}.
Certainly the massive fragment ``2'' is separated at the first step, then occurs the rupture of the
second neck between fragments  ``1'' and ``3''.
We will discuss the probability of the rupture at the second neck between ``1'' and ``3'', which
decreases with increasing the distance $R_{32}$ that induces the decrease of the Coulomb field
of the massive fragment ``2''.
The definitions of the variables of TNS used in this analysis can be found in Fig.~\ref{ThreeSph}.

\section{Theoretical approach: from a dinuclear system to a tri-nuclear system}

In order to explore the mechanisms of the CCT process, we apply the theoretical framework
of the dinuclear system (DNS) model~\cite{ACNPV93,ACNPV95,Volkov99,NFTA05}.
In the present work, we estimate the total energy of the interacting system by calculate the sum of
the binding energies of its constituents and the interaction potential energy between them.
The minima of the potential energy surfaces (PES) are found by the variation of the charge and mass
numbers of two fragments out of the three fragments and the distances between them. 
The PES is the two dimensional driving potential which depends on the charge numbers of 
two fragments of the collinear TNS. The distances $R_{13}$ and $R_{32}$ between centers of mass
of fragments are found from the minimum value of the nucleus-nucleus interaction. 
 
The fission process is considered as a formation of the elongated mononucleus (for example,
superdeformed shape) which breaks down into two fragments as in the case of binary fission.
The formation of the third cluster in the neck region and the splitting of this system
into three fragments are related to the shape of the system such as hyperdeformation.
Furthermore, the assumption of the formation of a heavier nucleus as the third fragment
between the two main fission products is also introduced.

\subsection{Total potential energy of a tri-nuclear system}

The study on the landscape of the potential energy surface (PES) is carried out to find minima and
valleys, since, at local maxima, one can expect increased yields of the mass and charge distributions
in the TNS undergoing fission process.
It should be noted that the stage of transition from compound nucleus to the TNS  configuration is not
analyzed.
Instead, we assume that TNS is formed during fission of the compound nucleus into a binary system.
This process can occur in the sense of energy conservation.
We refer to Ref.~\cite{CKT05} for the hyperdeformed \nuclide[236]{U} nucleus.

The PES is calculated as
\begin{eqnarray}
&& U(Z_1,A_1,\beta^{(1)},Z_2,A_2,\beta^{(2)},Z_3,A_3,\beta^{(3)},R_{13},R_{32})
\nonumber\\
&=& V_{\rm int}(Z_1,A_1,\beta^{(1)},Z_2,A_2,\beta^{(2)},Z_3,A_3,\beta^{(3)},R_{13},R_{32})
\nonumber \\ && \mbox{}
+Q_{\rm ggg}(Z_1,A_1,Z_3,A_3),
\label{Vtot}
\end{eqnarray}
where $Z_i$ and $A_i$ are the charge number and mass number of the $i^{\rm th}$
fragment of the TNS ($i=1,2,3$), respectively, and $R_{ij}$ is the distance between the
mass-centers of the $i^{\rm th}$ and $j^{\rm th}$ fragments.
Here, $\beta^{(i)}=\{ \beta_2^{(i)}, \beta_3^{(i)} \}$ is a set of deformation parameters
of the fragment $i$, where $\beta_2^{(i)}$ and $\beta_3^{(i)}$ represent the quadrupole
and octupole parts, respectively.
The interaction potential $V_{\rm int}$  between the fragments of TNS
can be written as
\begin{eqnarray}
&& V_{\rm int}(Z_1,A_1,\beta^{(1)},\beta^{(2)}, Z_3,A_3,\beta^{(3)},R_{13},R_{32})
\nonumber \\ &=&
\sum^{3}_{i<j}V_{ij}(Z_i,A_i,\beta^{(i)},Z_j,A_j,\beta^{(j)};R_{ij}),
\label{Vint}
\end{eqnarray}
where $V_{ij}$ is the two-body interaction potential between the nuclei ``$i$'' and ``$j$''.
It contains two parts, namely, the nuclear part $V^{ij}_{\rm nuc}$ and the Coulomb part $V^{ij}_C$,
so that
\begin{eqnarray}
&& V_{ij}(Z_i,A_i,\beta^{(i)},Z_j,A_j,\beta^{(j)};R_{ij})
\nonumber \\ &=&
V^{ij}_{\rm nuc}(Z_i,A_i,\beta^{(i)},Z_j,A_j,\beta^{(j)};R_{ij})
\nonumber\\ && \mbox{}
+V^{ij}_C(Z_i,A_i,\beta^{(i)},Z_j,A_j,\beta^{(j)};R_{ij}).
\label{Vij}
\end{eqnarray}
It is clear that $V^{12}_{\rm nuc} = 0$ since the fragments ``1'' and ``2'' are separated by the
fragment ``3'' and, therefore, there is no overlap of their nucleon densities.
The nuclear part of the nucleus-nucleus interaction $V^{ij}_{\rm nuc}$ is calculated  by using
the double folding procedure~\cite{TNS11}, and the Coulomb part $V^{ij}_C$ is estimated by
the Wong expression~\cite{Wong73}.

In Eq.~(\ref{Vtot}), $Q_{\rm ggg}$ is the reaction balance energy in ternary fission, which is
written as
\begin{eqnarray}
&& Q_{\rm ggg}(Z_1,Z_3,A_1,A_3)
\nonumber \\
&=& B_1(Z_1,A_1) + B_2(Z_2,A_2) + B_3(Z_3,A_3)
\nonumber\\ && \mbox{}
-B_{CN}(Z_{CN},A_{CN}).
\label{Qgg}
\end{eqnarray}
The values of binding energies $B_i$ for ground states are taken from Refs.~\cite{AWT03,MNMS93}.

In order to calculate the mass and charge distributions of the TNS in the pre-scission state,
the minima and valleys of the PES are determined by computing the interaction potential $V_{\rm int}$
as a function of $(Z_1, A_1, Z_3, A_3, R_{13}, R_{32})$ since $(Z_2,A_2)$ can be defined through
$(Z_1, A_1, Z_3, A_3)$ and $R_{12}=R_{13}+R_{32}$.
This is done by taking $V_{\rm int}$ as a function of  $R_{13}$ and $R_{32}$ for each configuration
of $\{Z_1,A_1; Z_3,A_3; Z_2,A_2\}$. (See Fig.~\ref{ThreeSph} for the geometry.)

In order to find the dominant cluster states of PES, the charge (and mass) numbers of the two
fragments are varied in the range of $2 < Z_1 < Z_{\rm CN}/2$ and
$2 < Z_3  < Z_{\rm CN}/2$ [$A_{1,\rm min} < A_1(Z_1) < A_{1,\rm max}$ and
$A_{3,\rm min} < A_3(Z_3) < A_{3,\rm max}$].
The charge and mass numbers of the third fragment can be found from the corresponding conservation
laws for them.
The distances $R_{13}$ and $R_{32}$ between interacting nuclei are then varied to find
$R^{(\rm min)}_{13}$ and $R^{(\rm min)}_{32}$ that correspond to  the minimum values of the potential
wells $V_{13}$ and $V_{32}$, respectively.
It should be noted again that the potentials are affected by the Coulomb interaction
$V^{C}_{12}$ of the border fragments.

This process allows us to find the mass number $A_i$ that corresponds to the minimum value of
the PES for a given value of $Z_i$.
For example, the value of $A_1$ can be found by minimizing the PES for each value of $A_3$
at fixed values of $Z_1$ and $Z_3$.
From the set of the results calculated for PES as a function of
$(Z_1, A_1, Z_3, A_3, R_{13}, R_{32})$ we can establish the driving potential demonstrating the
configurations of the TNS with the well-pronounced cluster states having closed shells.
The three-dimensional driving potential $U_{\rm dr}(Z_1, A_1; Z_3, A_3)$ is determined by the
values of the PES in Eq.~(\ref{Vtot}) corresponding to the minimum values of the potential wells in the
nucleus-nucleus interaction $V_{\rm int}$ between neighbor fragments as a function of the distances
between their centers-of-mass:
 \begin{eqnarray}
&& U_{\rm dr}(Z_1, A_1, \beta^{(1)},\beta^{(2)},Z_3, A_3, \beta^{(3)})
\nonumber \\ &=&
U(Z_1, A_1, \beta^{(1)},\beta^{(2)},Z_3, A_3, \beta^{(3)},R^{(\rm min)}_{13},R^{(\rm min)}_{32}).
\label{Udr}
 \end{eqnarray}
A change of $A_i$  leads to the change of $Q_{\rm ggg}$ which depends on the binding
energies $B_i$.
As a result, $U_{\rm dr}$ is sensitive to the mass distribution between the TNS fragments.

\subsection{Probability of the yield of ternary fission fragments}

The mass and charge distributions of the TNS fragments are related to the driving potential
$U_{\rm dr}$.
Therefore, the knowledge of $U_{\rm dr}$ allows us to calculate the yield of the products of
ternary fission as in Ref.~\cite{TNS11}:
\begin{eqnarray}
Y(Z_1,A_1; Z_3, A_3) &=& P(Z_1,A_1; Z_3, A_3)
\nonumber \\ && \mbox{} \times
W_{13}(Z_1,A_1; Z_3, A_3)
\nonumber \\ && \mbox{} \times
W_{32}(Z_1,A_1; Z_3, A_3),
\label{Yield}
\end{eqnarray}
where $P(Z_1,A_1; Z_3, A_3)$ is the probability of the charge and mass distributions of the
TNS fragments.
The probability of the formation of a TNS, $P(Z_1,A_1; Z_3, A_3)$, can be found from the
condition of a statistical equilibrium as in Ref.~\cite{MS75},
i.e., the TNS has an equilibrium state before scission:
\begin{widetext}
\begin{equation}
P(Z_1,A_1; Z_3, A_3)=
P_{0}\exp\left[-U_{\rm dr}(Z_1,A_1; Z_3, A_3)/T_{\rm TNS}(Z_1,A_1; Z_3, A_3)\right],
\label{Pz}
\end{equation}
\end{widetext}
where  $T_{\rm TNS}$ is the effective temperature of the TNS and
$U_{\rm dr}(Z_1,A_1; Z_3, A_3)$ is determined by the formula (\ref{Udr}).
The normalization coefficient for the yield probability is represented by $P_{0}$.

In Eq.~(\ref{Yield}), $W_{13}$ and $W_{32}$ are the decay probabilities of the TNS that are caused
by overcoming the pre-scission barriers $B_{13}$ and $B_{32}$ which correspond
to the separation of the first and second nuclei, respectively.
Their explicit expressions can be found as~\cite{TNS11}
\begin{eqnarray}
\label{W13}
W_{13}(Z_1,A_1; Z_3, A_3)&=&W^{0}_{13}\exp\left[-\frac{B_{13}}{T_{13}}\right],\\
W_{32}(Z_2,A_2; Z_3, A_3)&=&W^{0}_{32}\exp\left[-\frac{B_{32}}{T_{32}}\right],
\label{W32}
\end{eqnarray}
where  $(B_{13},B_{32})$ and $(T_{13}, T_{32})$ are the pre-scission barriers and the effective temperatures on these barriers the corresponding parts of the TNS.
The barriers $B_{13}$ and $B_{32}$ prevent the separation of the outer fragments from the middle one.
These pre-scission barriers are defined by the depth of the nucleus-nucleus potential well
between neighbor fragments of the TNS.
Here, $W^{0}_{13}$  and $W^{0}_{32}$ are normalization coefficients for the corresponding
probability distributions.

The effective temperatures are determined by the excitation energy of the TNS which is generated by
the descent of the system from the saddle point in binary fission.
We assume that the third cluster is formed between the two parts of fissioning nuclei before their
splitting.
In this case, $E^*_{\rm TNS}(Z_1, A_1, Z_3, A_3)$, the excitation energy of the system,
is determined by the difference between the values at the saddle point and at the point of
the minimum driving potential with the considered charge and mass numbers of clusters:
\begin{equation}
E^*_{\rm TNS}(Z_1, A_1, Z_3, A_3)=E^*_{\rm CN}-U_{\rm dr}(Z_1,A_1;Z_3, A_3).
\end{equation}

The effective temperatures of the TNS,  necks ``1-3'' and ``2-3''  are defined by the excitation
energies on the minimum of the driving potential and pre-scission barriers $B_{13}$ and $B_{32}$, respectively:
\begin{eqnarray}
T_{\rm TNS}^{} &=& \sqrt{\frac{12E^*_{\rm TNS}}{A_{CN}}},
\nonumber  \\
T_{13} &=& \sqrt{\frac{12E^*_{13}}{(A_{1}+A_3)}} ,
\nonumber \\
T_{32} &=&\sqrt{\frac{12E^*_{32}}{(A_2+A_3)}},
\end{eqnarray}
where $E^*_{13}$ and $E^*_{32}$ are the excitation energies on the top of the pre-scission
 barrier the DNS ``1-3'' and of the DNS ``2-3'', respectively. These excitation energies
are the result of sharing the TNS excitation energy between different degrees of freedom.
The parts of $E^*_{\rm TNS}$ corresponding to the decay axes  $R_{13}$ and $R_{32}$ 
are estimated by assumption that their inertia masses are $A_{13}=A_1 (A_2+A_3)/A_{\rm CN}$ and
$A_{32}=A_2 (A_1+A_3)/A_{\rm CN}$, respectively. Then the values of $E^*_{13}$ and $E^*_{32}$ 
are found from the effective temperature of the TNS:
\begin{equation}
E^*_{\rm 13}(Z_1, A_1, Z_3, A_3)=\frac{T_{\rm TNS}^2 A_{13}}{12}-B_{13}, \\
\end{equation}
\begin{equation}
E^*_{\rm 32}(Z_1, A_1, Z_3, A_3)=\frac{T_{\rm TNS}^2 A_{32}}{12}-B_{32}.
\end{equation}
If the residual part of the TNS excitation energy 
$E^*_{\rm res}=E^*_{\rm TNS}-E^*_{\rm 13}-E^*_{\rm 32}$ is larger than the energy $B_n$ 
 for the emission of neutrons from the TNS fragments, 
 the ternary fission is accompanied by neutrons.
 

\section{\boldmath Investigation of tripartition in the \nuclide[252]{C\lowercase{f(sf)}} reaction}
\label{tripartition}

\subsection{Potential energy surface showing the cluster formation in TNS}
\label{Population}

In the experiment reported in Ref.~\cite{PKVA12}, the ternary products were formed in the
spontaneous fission of \nuclide[252]{Cf} and the yields of \nuclide[68]{Ni},
\nuclide[80-82]{Ge}, \nuclide[94]{Kr},  \nuclide[128,132]{Sn}, and \nuclide[144]{Ba} were obtained.
In the plot of the mass-mass distribution of two products (third one is missing) given in Fig.~10
of Ref.~\cite{PKVA12}, these events were found to form a rectangle, and the authors of
Ref.~\cite{PKVA12} assumed that the points in the right half of the rectangle likely
reflect the shell effects around $N=88$.

\begin{figure*}[t]
\begin{center}
\includegraphics[width=1.20\textwidth]{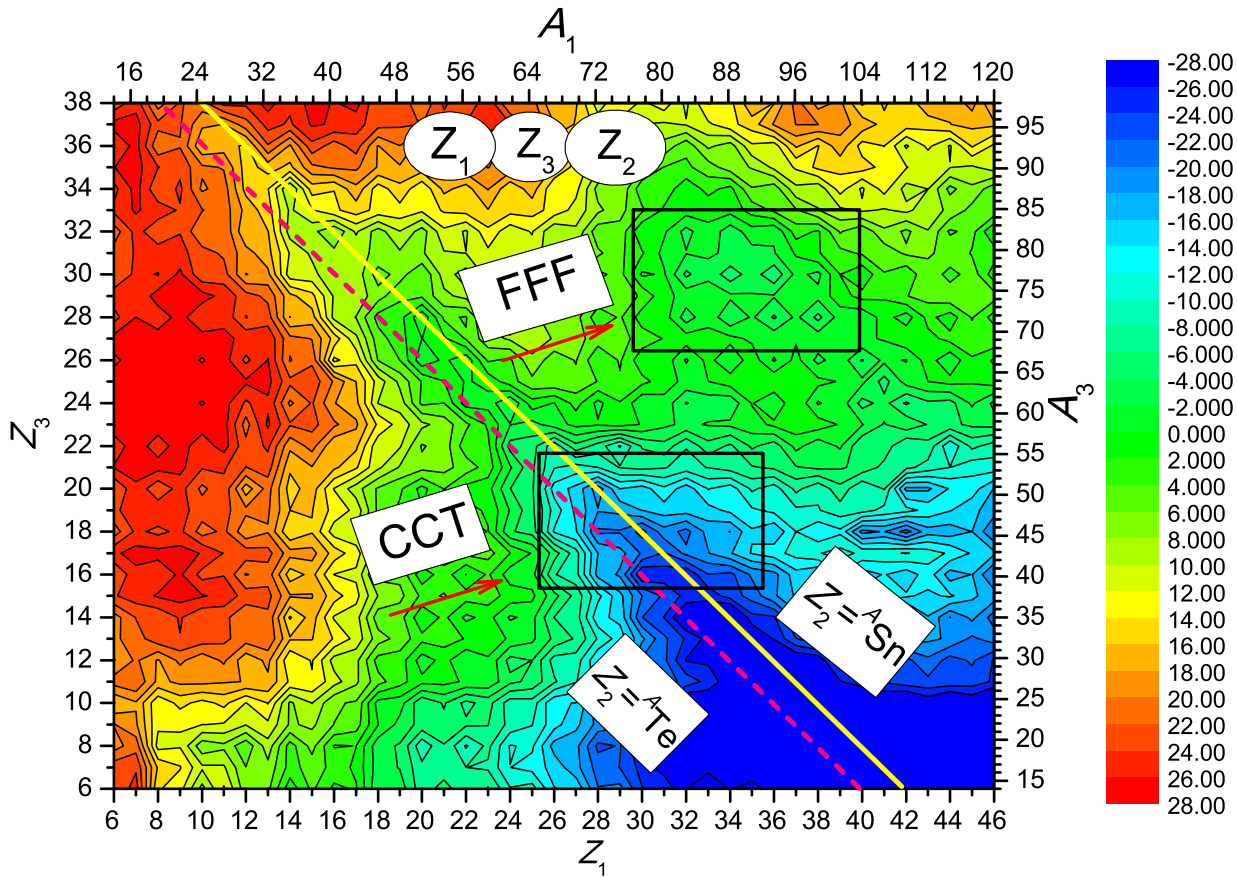}
\end{center}
\vspace*{-3.5 cm}
\caption{(Color online)
The potential energy surface of the \nuclide[252]{Cf(sf,fff)} reaction. The rectangle ``CCT''
 shows the area of the mass numbers $Z_1 (A_1)$ and $Z_3 (A_3)$ which corresponds to the CCT products. The rectangle ``FFF''  shows the area of formation three fragments with the
close mass numbers.  The solid and dashed lines show the TNS configuration having
$^{132}$Sn and $^{134}$Te, respectively, as the outer nucleus $Z_2$.}
\label{PES}
\end{figure*}

\begin{figure*}[t]
\begin{center}
\includegraphics[width=0.6\textwidth]{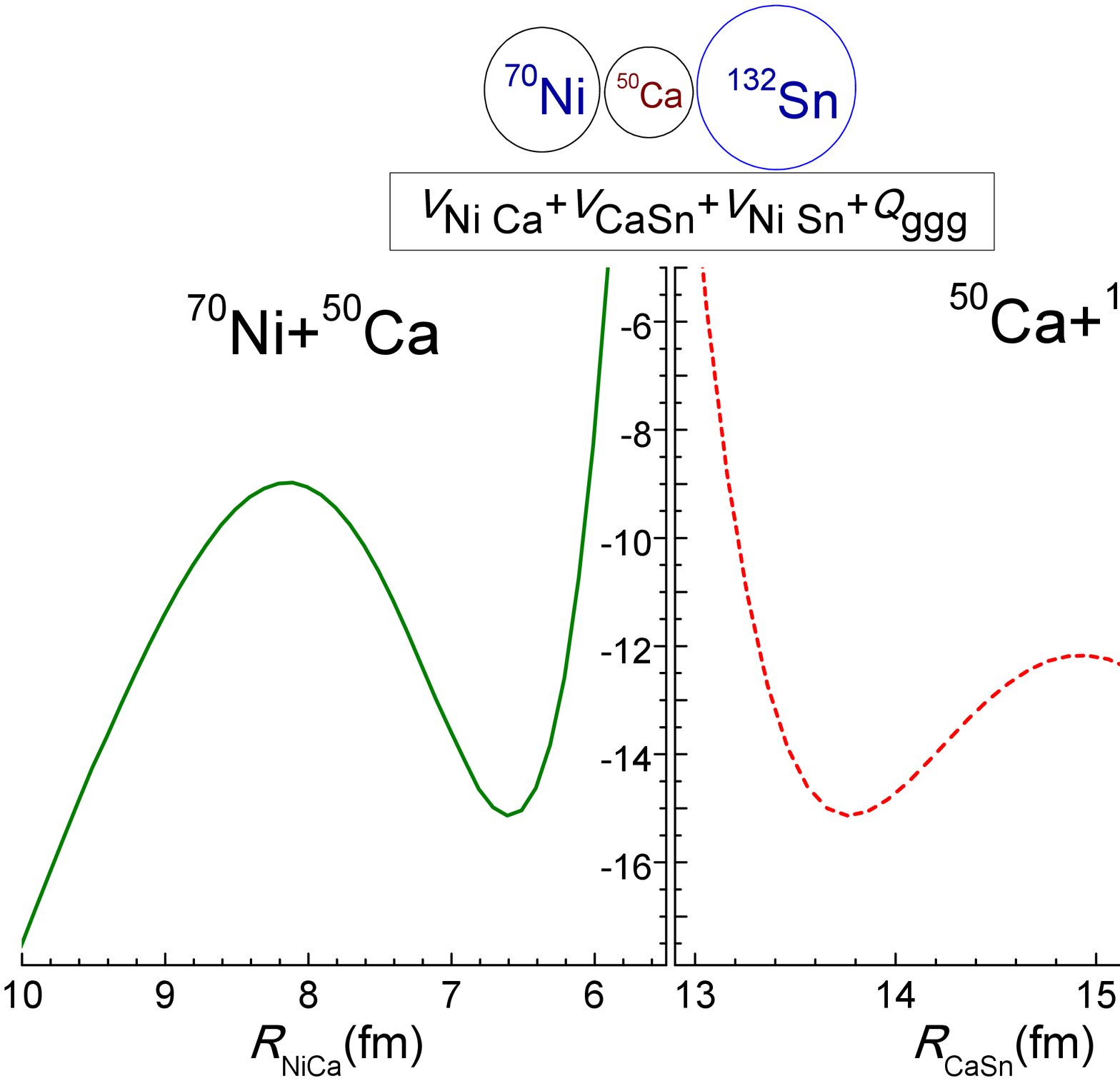}
\end{center}
\vspace*{-3.0 cm}
\caption{(Color online)
The pre-scission barriers $B_{\nuclide[]{Ni} \nuclide[]{Ca}}$ and $B_{\nuclide[]{Ca} \nuclide[]{Sn}}$
keeping TNS fragments together.}
\label{CaNiBarr}
\end{figure*}

The effect of the shell structure of the proton and neutron single-particle states 
in the formation of a
tri-nuclear system and in its decay into the observed fission products is obviously seen in the
mass-mass distribution data of Ref.~\cite{PKVA12}.
This observation stimulates us to calculate PES, i.e., $V(Z_1,Z_3,A_1,A_3,R_{13},R_{32})$,
and the driving potential $U_{\rm dr}(Z_1,A_1,Z_3,A_3)$ for the intermediate system preceding to
their formation.
The products of a CCT-decay should be formed before being separated from the other part of the
system.

\begin{figure*}[t]
\begin{center}
\includegraphics[width=0.8\textwidth]{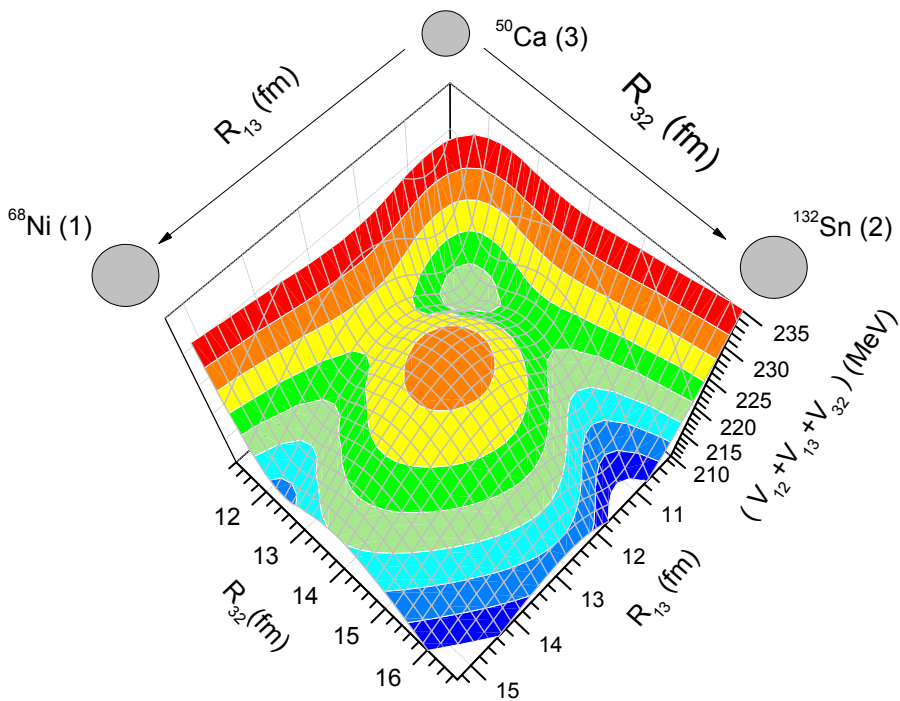}
\end{center}
\vspace*{-5.0 cm}
\caption{(Color online)
The total nucleus-nucleus interaction potential
$V_{\rm int}$ as a function of inter-center distances $R_{13}$
and $R_{32}$ between fragments of the TNS with collinear geometry.}
\label{3DV1V2}
\end{figure*}

Our results for the PES are presented in Fig.~\ref{PES}.
Each point in the driving potential $U_{\rm dr}(Z_1,A_1,Z_3,A_3)$ for the TNS corresponds to a
specific configuration (channel), which consists of three interacting nuclei placed in one line as shown in Fig.~\ref{ThreeSph}.
In calculation of PES, the distances between the fragments are fixed at the values
  corresponding  to the minimum values of the corresponding wells in the
 interaction potential between them  (see Figs. \ref{CaNiBarr} and \ref{3DV1V2}).
      The quadrupole deformation parameters  of the  first-excited $2^+$
    state of nuclei \cite{Raman} are used in calculation of PES.

The rectangle ``CCT''  shows the area of the mass numbers $Z_1 (A_1)$ and $Z_3 (A_3)$ which corresponds to the CCT products. The rectangle ``FFF''  shows the area of formation three fragments with the close mass numbers.  The solid and dashed lines show the TNS configuration having
$^{132}$Sn and $^{134}$Te, respectively, as the outer nucleus $Z_2$.
So, we can see the valley, which is the energetically minimum area ($Z_2=50$ and $Z_2=52$)
and corresponds to the $\nuclide[252]{Cf} \to \nuclide[]{f}_1 + \nuclide[]{f}_3 + \nuclide[132]{Sn}$ fission channel. The valley extends up to the area about $Z_3 = 28$.
As was mentioned earlier, $Z_3$ is the charge number of the middle cluster.
The phase space of the configurations corresponding to the blue color region is large.
Therefore, the probability of finding the TNS of configurations with a lower potential energy
is larger.
The configuration of $\nuclide[]{Ni} + \nuclide[]{Ca} + \nuclide[]{Sn}$ has large probability
compared with the $\nuclide[]{Ca} + \nuclide[]{Ni} + \nuclide[]{Sn}$ configuration since the PES value
of the latter configuration is about 12~MeV higher than that of the former configuration.

The calculations were performed to find the yield of the CCT products from the collinear geometry
based on the formula of Eq.~(\ref{Yield}).
In these calculations we found that the value of the pre-scission barrier plays the decisive role.
Therefore, in the next section we discuss the behavior of the barriers $B_{13}$ and $B_{32}$ for the
CCT channel of the $\nuclide[]{Ni} + \nuclide[]{Ca} + \nuclide[]{Sn}$ configuration.

\subsection{The decrease of the pre-scission barrier due to the Coulomb field of outer fragments}
\label{barrier}

The mechanism of almost sequential ruptures of the two necks connecting the fragments
of a collinear TNS is suggested to explain the observed yields of heavy clusters such as
\nuclide[]{Ni}, \nuclide[]{Ge}, and \nuclide[]{Se} isotopes that appear with the product having
a mass number of $A = 138$--$148$ in the CCT of \nuclide[252]{Cf}~\cite{PKVA10,PKVA12}.
The PES shows the structure of valleys and local minima that correspond to the formation of heavy
clusters observed in experiments as shown in Fig.~\ref{PES}.
These fragments of a TNS should be emitted from the potential wells, and, therefore,
it is important to estimate the depths of the potential wells, since heavy clusters are allowed to exist during a definite long time.
In Fig.~\ref{CaNiBarr} the potential wells calculated for the TNS of
$\nuclide[]{Ni} + \nuclide[]{Ca} + \nuclide[]{Sn}$, which forms a linear chain, are presented
as functions of the distances between centers of the middle nucleus
(\nuclide[]{Ca}) and the outer nuclei (\nuclide[]{Ni} and \nuclide[]{Sn}).
The values of these nucleus-nucleus potentials  are shifted by the values of $Q_{\rm ggg}$ 
 as the contour plot of the PES [see Fig.~\ref{PES} 
Eq. (\ref{Vtot})] to take into account the change of the intrinsic energy of the TNS.

\begin{figure}[t]
\begin{center}
\includegraphics[width=0.6\textwidth,clip=]{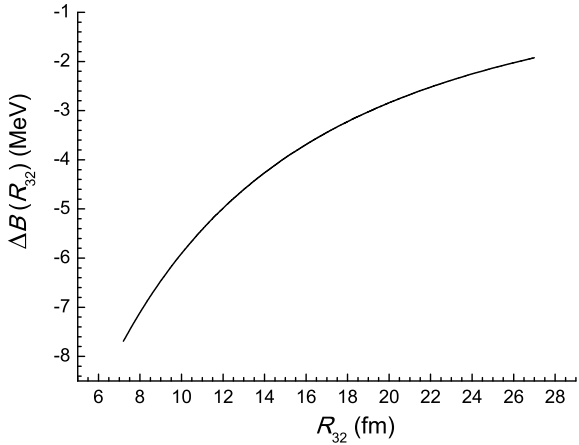}
\end{center}
\vspace*{-2 cm}
\caption{
The dependence of the pre-scission barrier $B_{13}$ for the decay of the binary system
$\nuclide[]{Ni} + \nuclide[]{Ca}$ on the distance $R_{32}$ due to the Coulomb interaction
of the massive third fragment \nuclide[]{Sn} in the collinear geometry.}
\label{Dwell}
\end{figure}

\begin{figure*}[t]
\begin{center}
\vspace*{3cm}
\includegraphics[width=1.0\textwidth,clip=]{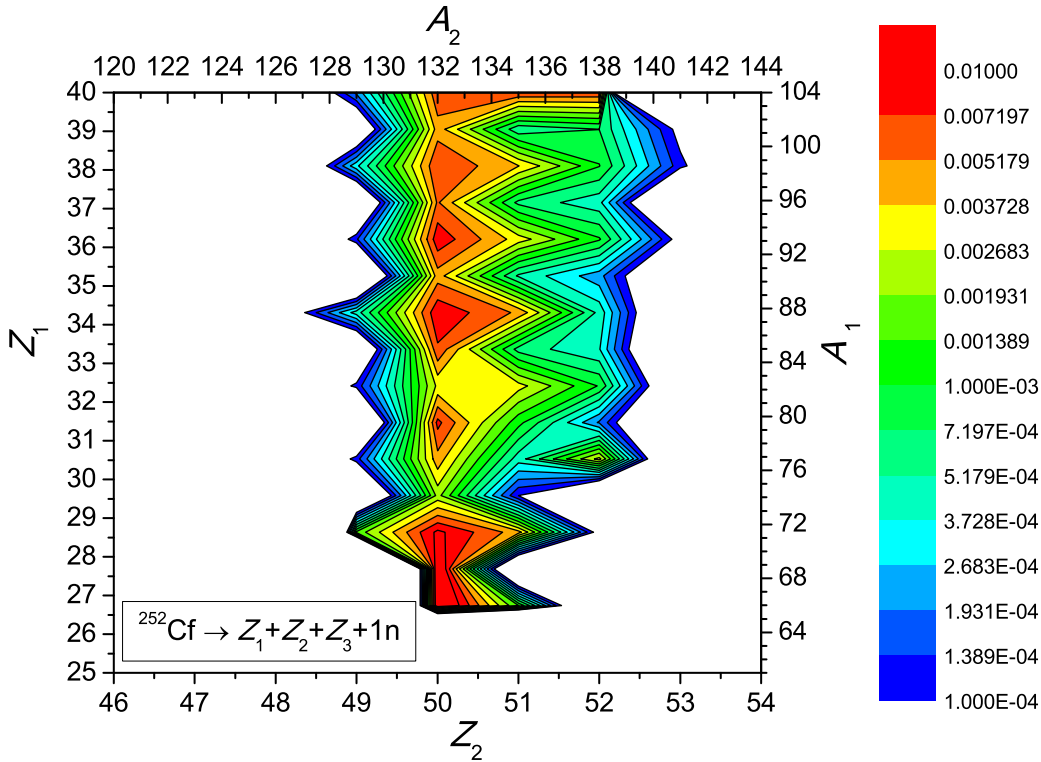}
\end{center}
\vspace*{-3.3cm}
\caption{(Color online)
Theoretical results for the yield of the outer fragments \nuclide[A_1]{Z_1} and \nuclide[A_2]{Z_2} of
the TNS formed at the spontaneous fission of \nuclide[252]{Cf}.}
\label{YZ1Z2}
\end{figure*}

For the calculation of the interaction potential $V_{\nuclide[]{Ca} \nuclide[]{Sn}}$,
the distance $R_{\nuclide[]{Ni} \nuclide[]{Ca}}$ is fixed to the value corresponding
to the minimum of $V_{\nuclide[]{Ni} \nuclide[]{Ca}}$, while $V_{\nuclide[]{Ni} \nuclide[]{Ca}}$
potential is calculated at the fixed value of $R_{\nuclide[]{Ca} \nuclide[]{Sn}}$ that gives the
minimum of $V_{\nuclide[]{Ca} \nuclide[]{Sn}}$.
The results for the nucleus-nucleus interaction between the nuclei of the collinear TNS of
$\nuclide[]{Ni} + \nuclide[]{Ca} + \nuclide[]{Sn}$ as a function
of the independent variables  $R_{13}$ ($R_{\nuclide[]{Ni} \nuclide[]{Ca}}$) and
$R_{32}$ ($R_{\nuclide[]{Ca} \nuclide[]{Sn}}$) are given by a three dimensional plot of the PES
in Fig.~\ref{3DV1V2}.
The contour lines of the PES presented in Fig.~\ref{PES} is calculated with the minimum value of the
nucleus-nucleus interaction at $R_{13}=11$~fm and $R_{32}=12$~fm in Fig.~\ref{3DV1V2}.
The decay of the TNS occurs due to the motion of the system along  $R_{13}$ or $R_{32}$.
The height of the pre-scission barrier is smaller in the direction along $R_{32}$
($R_{\nuclide[]{Ca} \nuclide[]{Sn}}$ in Fig.~\ref{CaNiBarr}) and, therefore, the massive fragment
\nuclide[]{Sn} is separated first from the TNS. This result is obtained by the use of Eqs. (\ref{W13}) and (\ref{W32}).
If the residual $\nuclide[]{Ni} + \nuclide[]{Ca}$ part of the TNS does not decay, the binary decay would
be observed since the $\nuclide[]{Ni} + \nuclide[]{Ca}$ system is considered as a superdeformed
shape of \nuclide[48]{Cd}.

\begin{figure*}[t]
\begin{center}
\vspace*{3cm}
\includegraphics[width=1.0\textwidth,clip=]{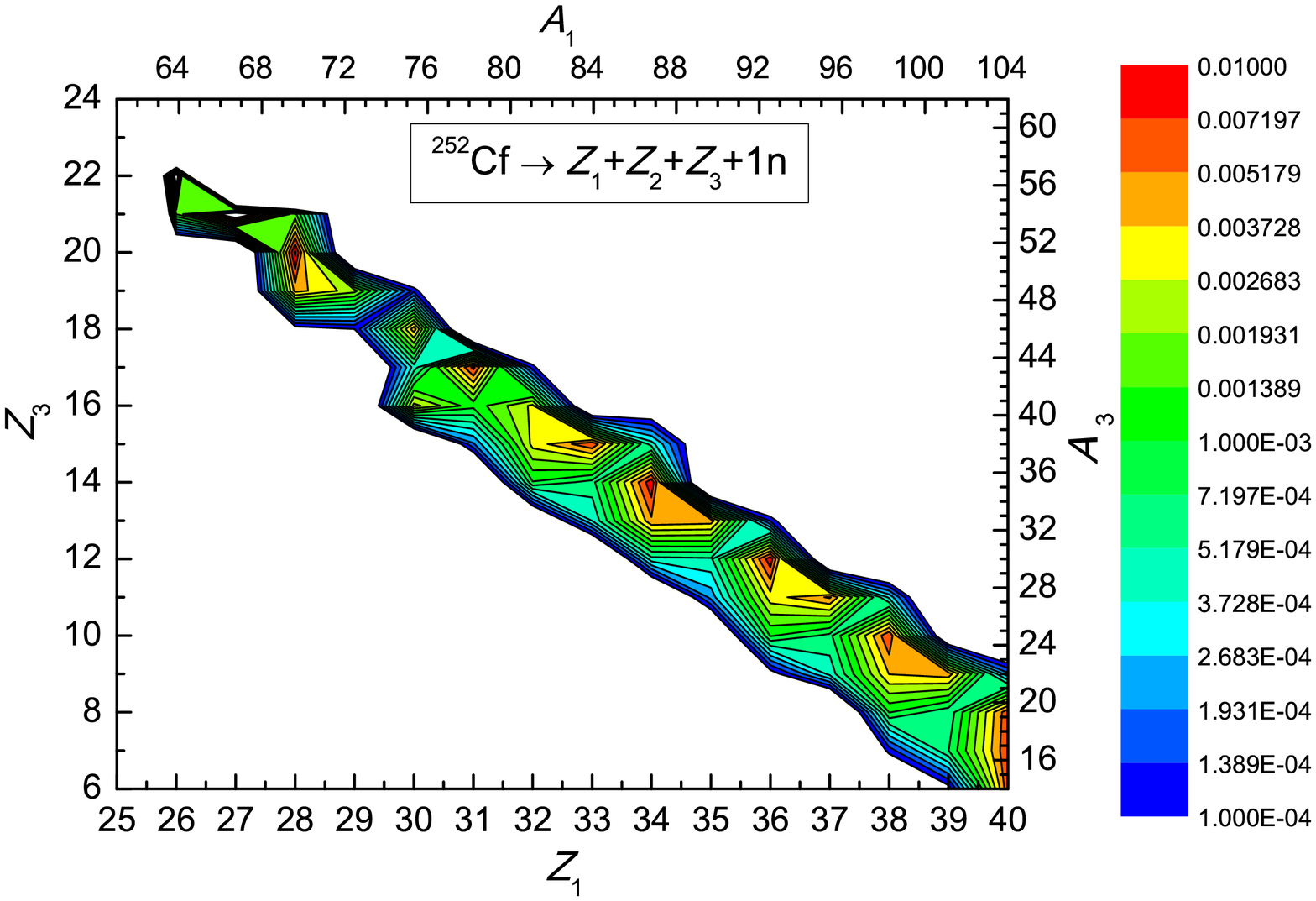}
\end{center}
\vspace*{-2.8 cm}
\caption{(Color online)
Theoretical results for the yield of the outer \nuclide[A_1]{Z_1} and middle \nuclide[A_3]{Z_3} fragments
of the TNS formed at the spontaneous fission of \nuclide[252]{Cf}.}
\label{YZ1Z3}
\end{figure*}

The excitation energy of the residual $\nuclide[]{Ni} + \nuclide[]{Ca}$ system should be large enough
so that it decays into \nuclide[]{Ni} and \nuclide[]{Ca}, if these nuclei are observed as CCT products.
The probability of this event strongly depends on the position of the separated massive product, i.e.,
the \nuclide[]{Sn} nucleus.
The depth of the potential $V_{\nuclide[]{Ni} \nuclide[]{Sn}}$, which is the pre-scission barrier
$B_{\nuclide[]{Ni} \nuclide[]{Sn}}$, changes as a function of the distance $R_{32}$.
To show this phenomenon we estimate the change of the $B_{13}$ barrier, which is the difference
between the maximum (on the barrier) and the minimum values of $V_{\rm int}$ as a function of
$R_{13}$ in Eq.~(\ref{Vint}).
The dependence of the change of the barrier $B_{13}$  by the distance $R_{32}$ is reduced to a
simple form of
\begin{equation}
\Delta B_{13}(R_{32})=\frac{Z_1 Z_2 e^2}{(R^{(B)}_{13}+R_{32})}
-\frac{Z_1 Z_2 e^2}{(R^{\rm (min)}_{13}+R_{32})},
\end{equation}
where $e^2 = 1.44$~MeV$\cdot$fm.
The dependence of $B_{13}$ on $R_{32}$ is presented in Fig.~\ref{Dwell}.
The negative values mean the decrease of the depth of the potential well  
($B_{13}(R_{32} \rightarrow \infty)+\Delta B_{13}(R_{32})$) in the interaction 
of the $\nuclide[]{Ni} + \nuclide[]{Ca}$ system.
The main observation of the present work is that the presence of the third fragment is important to
cause the decay of the $\nuclide[]{Ni} + \nuclide[]{Ca}$ system in an easier way.
The presence of the third massive fragment \nuclide[]{Sn} makes the pre-scission barrier shallower
by $4$~MeV, and thus the decay probability of the $\nuclide[]{Ni} + \nuclide[]{Ca}$ system increases.

By taking into account the change of the pre-scission barrier one can obtain reasonable
results for the yields of the \nuclide[]{Ni} isotopes followed by the emission
 of massive \nuclide[]{Sn} isotopes from the formula in
Eq.~(\ref{Yield}) that includes the effects of the pre-scission barriers $B_{13}$ and $B_{32}$.
The results are presented in Figs.~\ref{YZ1Z2} and \ref{YZ1Z3}.
In the former figure we use $Z_1$ ($A_1$) and $Z_2$ ($A_2$) axes for the plot, while in the latter
figure we use $Z_1$ ($A_1$) and $Z_3$ ($A_3$) axes.

Although the calculated yields of heavy clusters such as \nuclide[]{Ni}, \nuclide[]{Ge}, and
\nuclide[]{Se} isotopes are found to be in good agreement with the experimental data, there still
remains a difference between the mass numbers of the massive CCT products of \nuclide[252]{Cf}
observed in Refs.~\cite{PKVA10,PKVA12}, namely $A = 138$--$148$ which overlap with our results with
$A = 132$--$140$ presented in Figs.~\ref{YZ1Z2} and \ref{YZ1Z3}.
The strong yield of the products with mass numbers $A=132$--$140$ indicates the important role
of the magic number of neutrons at $N=82$.
This difference may be ascribed to that we use the tabulated masses of Refs.~\cite{AWT03,MNMS93}
to obtain the binding energies of nuclei.
This procedure, in fact, gives the binding energies of the ground states, but we may have deformed
nuclei at the scission point, which is highly probable for massive nuclei. But we should remind 
the procedure of calculation of the PES by variation of the charge and mass numbers 
of the two fragments of TNS ($2 < Z_1 < Z_{\rm CN}/2$ and
$2 < Z_3  < Z_{\rm CN}/2$ [$A_{1,\rm min} < A_1(Z_1) < A_{1,\rm max}$ and
$A_{3,\rm min} < A_3(Z_3) < A_{3,\rm max}$]). The depends of shell corrections  on the 
deformation should be studied for the most of the numerous (some thousands) combinations. 
Since the primary goal of this work is to demonstrate the possibility of the formation of the \nuclide[]{Ni},
\nuclide[]{Ge}, and \nuclide[]{Se} isotopes and their yields in the CCT mechanism, we leave more
accurate and sophisticated description of the production of massive isotopes of $A = 138$--$148$ to
our future studies.


\section{Estimation of the kinetic energy of the middle fragment}

The range of the kinetic energy of the middle fragment ``3'' can be estimated by
applying the energy and momentum conservation laws.
For simplicity, we assume that the kinetic energy of the binary process is determined by the Coulomb
barrier of the nucleus-nucleus interaction.
The first step of the sequential collinear ternary fission is the separation of the right fragment ``2'' as
shown in Fig.~\ref{graph1}.
The sum of the kinetic energies of this fragment and the residual part of the TNS is the same as the
Coulomb repulsion between them, which leads to
\begin{eqnarray}
 \frac{Z_1 Z_2 e^2}{R_{13} + R_{23} + d} + \frac{Z_3 Z_2 e^2}{R_{23} + d}
&=& \frac{m (A_1+A_3) v^2_{13}}{2} \nonumber\\
&+& \frac{m A_2 v^2_{2}}{2} \, ,
 \\
 m (A_1+A_3) v_{13}^{} + m A_2 v_{2}^{} &=& 0 \, ,
\end{eqnarray}
where $v_{13}^{}$ and $v_{2}^{}$ are the relative velocities of the DNS ``13'' and of the separated
fragment ``2'', respectively, in the laboratory frame.
The free parameter $d$ is introduced to decrease the sum of the total Coulomb barriers that can not
be larger than the reaction energy balance $Q_{\rm ggg}$ given in Eq.~(\ref{Qgg}).
The second step of the sequential collinear ternary fission is a decay of the DNS ``13'' into two
fragments ``1'' and ``3''.
The sum of their kinetic energies is then equal to the Coulomb repulsion between them so that
\begin{eqnarray}
&& \frac{Z_1 Z_3 e^2}{R_{13}} = \frac{m A_1 v'^{2}_{1}}{2} + \frac{m A_3 v'^{2}_{3}}{2} \, ,
\nonumber \, \\
&& m A_1 v'_{1} + m A_3 v'_{3} = 0 \, ,
\end{eqnarray}
where $v'_{1}$ and $v'_{3}$ are the velocities of the fragments ``1'' and ``3'', respectively, in the
moving frame with velocity $v_{13}^{}$ in the opposite direction to $v_2^{}$.
Therefore, we have
\begin{eqnarray}
v_1^{} &=& v'_{1} + v_{13}^{}  \, ,
\nonumber \\
v_3^{} &=& -v'_{3} + v_{13}^{} \, .
\end{eqnarray}

\begin{figure*}[t]
\centering
\includegraphics[width=0.9\textwidth,clip=]{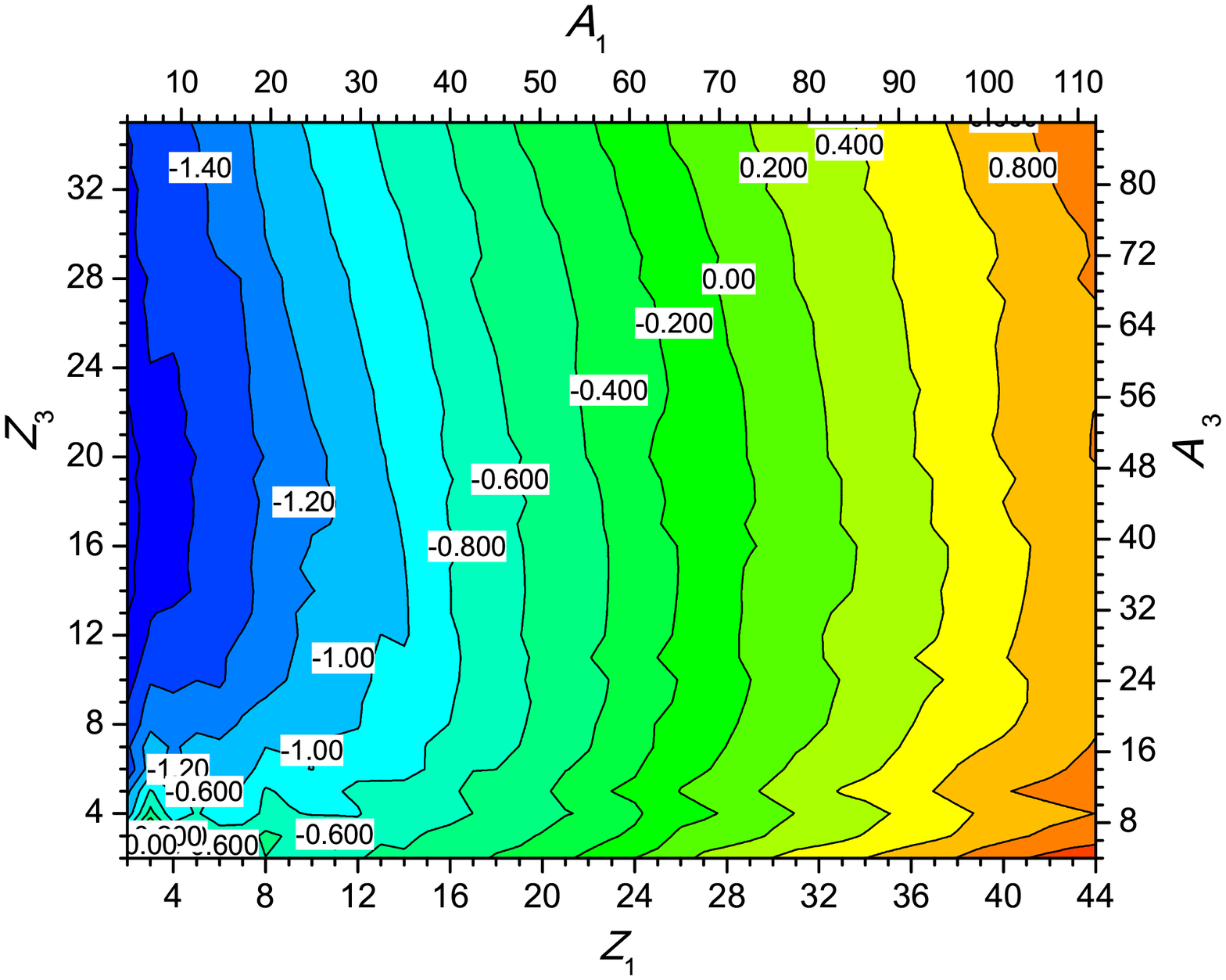}
\vspace*{-2.4 cm}
\caption{(Color online)
The contour map of the calculated velocity $v_3^{}$ (in cm/ns)
of the middle \nuclide[A_3]{Z_3} fragment of the TNS
formed at the spontaneous fission of \nuclide[252]{Cf} as a function of the mass numbers of the outer
fragments \nuclide[A_1]{Z_1} and \nuclide[A_2]{Z_2}. The negative values of $v_3^{}$
mean that its direction is opposite to the direction of $v_2$. }
\label{Fig9}
\end{figure*}

We then obtain $v_3^{}$ as a function of the mass numbers of the outer fragments \nuclide[A_1]{Z_1}
and \nuclide[A_2]{Z_2} and the results are presented in Fig.~\ref{Fig9}.
The negative values of $v_3^{}$ mean that its direction is opposite to the direction of $v_2$.
This figure also allows us to find the region of mass numbers $A_1$ and $A_2$ where the
velocity of the middle cluster is large enough to be registered in experiments.
One of the puzzles in the experimental data on the collinear tripartition presented in
Refs.~\cite{PKVA10,PKKA10,PKVA12} is the missing of the third fragment.
As can be understood from this analysis, the main physical reason of this phenomenon
is the smallness of the velocity of the ``missing" third product.

In Fig.~\ref{Fig9} one can see that the third product has a small velocity ($|v_3|<0.25$cm/ns) 
 for the case of
$A_1 = 60$--$80$ and $A_3 = 24$--$64$, which means that the range of mass numbers
 for the massive fragment is $A_2 = 108$--$168$.
This region overlaps with the observed mass region, where a \nuclide[]{Ni}-like product with a mass
number of $A_1 = 68$--$72$ was observed with a massive product of
$A_2 = 136$--$144$~\cite{PKVA10,PKKA10,PKVA12}.
In the case of the symmetric masses, $A_1 \sim A_2 \sim A_3$, we have a small velocity of the middle
fragment $A_3$, namely, we get $v_3^{} = 0.3$--$0.4$~cm/ns.
The range of the mass numbers where the third middle fragment has an observable velocity
is found to be $A_1 = 100$--$120$ and $A_3 = 4$--$16$ (i.e., $A_2 = 116$--$148$) which corresponds
to the well-known ternary fission with emission of the light nuclei with a mass number of
$A_3 = 4$--$12$~\cite{Goenn04,GMK05}.
In the experiments reported in Refs.~\cite{Goenn04,GMK05}, all the three ternary fission products
were registered.
The other range of mass numbers of the outer fragments of TNS which allows to the observation of
the middle fragment is $A_1 = 104$--$112$ and $A_3 = 64$--$90$ ($A_2 = 50$--$84$).
The decay channel of $A_2 < 100$ has a very small probability to be realized since the pre-scission
barrier $B_{32}$ is sufficiently high.
Our analysis on the sequential true ternary fission shows that the possibility of observing the middle
fragment in this case is small.


\section{Conclusion}

In this work, we suggested a sequential ternary fission process with a very short time between
the ruptures of two necks connecting the middle cluster of the collinear tri-nuclear system.
The necessity of this mechanism is revealed in the decrease of the pre-scission barrier of the
residual part of the TNS due to the Coulomb field of the massive fragment being separated first.
This mechanism leads to the probability of about $10^{-3}$ for the yield of massive clusters such
as \nuclide[70]{Ni},  \nuclide[80-82]{Ge}, \nuclide[86]{Se}, and \nuclide[94]{Kr} produced with
a product of $A = 132$--$140$ in the CCT of \nuclide[252]{Cf}.
The yields of  such products were observed in coincidence with a massive product of
$A = 138$--$148$ with a relatively large probability in the experiments of the FOBOS group at
the FLNR of JINR (Dubna).

To verify the realization of this mechanism, the total potential energy of the chain-like TNS was
calculated as a sum of $Q_{\rm ggg}$ and the nucleus-nucleus interaction potential energy between
its constituents.
The minima and valleys of the PES related to the shell effects in nuclei were determined by using the
binding energies obtained from the well-known mass tables of Refs.~\cite{AWT03,MNMS93} and the
calculation of the interaction potential for the charge and mass numbers of the three fragments as a
function of distances between their centers-of-mass.
The distances $R_{13}$ and $R_{32}$ between interacting nuclei were varied to find the minimum
values of the potential wells of $V_{13}$ and $V_{32}$, respectively, which are affected by the Coulomb
interaction $V^{C}_{12}$ of the border fragments.
The driving potential as a function of the charge and mass numbers of two fragments was obtained at
the values of the distances $R^{(\rm min)}_{13}$ and $R^{(\rm min)}_{32}$ that correspond to the
minimum values of the nucleus-nucleus interactions $V_{13}$ and $V_{32}$, respectively.

In order to find the dominant cluster states of the TNS, the driving potential
$U_{\rm dr}(Z_1,A_1,Z_3,A_3)$ was calculated for the values of the charge (mass)  numbers of the
two fragments in the ranges of $2 < Z_1 < Z_{\rm CN}/2$ and $2 < Z_3 < Z_{\rm CN}/2$
[$A_{1,\rm min} < A_1(Z_1) < A_{1,\rm max}$ and the $A_{3,\rm min} < A_3(Z_3) < A_{3,\rm max}$].
The analysis of the results allows us to find the mass number $A_i$ corresponding to the minimum
value of the PES for a given value of $Z_i$.
The calculated total potential energy as a function of $(Z_1, A_1, Z_3, A_3)$ enabled us to
establish the three dimensional driving potential that demonstrates the configurations of TNS with
probable cluster states in the pre-fission states.

Finally, the contour lines of the three dimensional driving potential showed the structure of a valley
corresponding to the formation of the outer cluster with $Z_2 = 50$ or $N_2 = 82$ at the ternary fission,
which corresponds to the fission channel of
$\nuclide[252]{Cf} \to \nuclide[]{f}_1 + \nuclide[]{f}_3 + \nuclide[132]{Sn}$.
It was found that the valley extends up to the area of about $Z_3 = 28$ and the probability of a
configuration having lower potential energy for the TNS is large.
Therefore, the configuration of $\nuclide[]{Ni} + \nuclide[]{Ca} + \nuclide[]{Sn}$ has a
large probability in comparison with the configuration of $\nuclide[]{Ca} + \nuclide[]{Ni} + \nuclide[]{Sn}$
since the PES value of the latter configuration is about 12~MeV higher than that of the former
configuration.

The dependence of the velocity of the middle cluster on the mass numbers $A_1$ and $A_2$ was
also analyzed for the case of the collinear tripartition of \nuclide[252]{Cf}.
The main physical reason associated with the collinear tripartition is the smallness of the missed
third product.
We found that, in the range of the mass numbers $A_1 = 60$--$80$ and $A_2 = 132$--$140$, the
middle fragment has a very small velocity, which is in agreement with the observed range
of values presented in ~Refs. \cite{PKVA10,PKKA10,PKVA12}.
This means that it is indeed difficult to observe the middle product of a collinear tripartition of
\nuclide[252]{Cf} producing \nuclide[]{Ni} with the second product having a mass number of
$A_2 = 132$--$140$.
In the case of the symmetric masses, $A_1 \sim A_2 \sim A_3$, we have a small velocity of the middle
fragment $A_3$, namely, we get $v_3^{} = 0.3$--$0.4$~cm/ns. The smallness of the middle cluster 
velocity may be a reason of its missing in the collinear tripartition in the   $^{252}$Cf(sf,fff) \cite{PKVA10} and $^{235}$U(n$_{\rm th}$,fff) \cite{PKVA12} reactions.

The mass ranges of the two outer products, where the middle fragment can be observed, are
i) $A_1 = 100$--$120$ and  $A_2 = 130$--$140$ which corresponds to the well-known ternary fission
with an emission of the light nucleus with $A_3 = 4$--$12$~\cite{Goenn04,GMK05}
and ii) $A_1 = 90$--$110$ and  $A_2 = 100$--$132$.

\acknowledgments

We thank our colleagues, Prof. D. V. Kamanin and Prof. Yu. V. Pyatkov, for fruitful discussions.
A.K.N. is grateful to RFBR for partial support and W.v.O. thanks FLNR of JINR for their hospitality
during his stay in Dubna.
\newblock
A.K.N. was supported in part by the MSIP of the Korean Government under the Brain Pool
Program No. 142S-1-3-1034.
\newblock
The work of Y.O. was supported by the National Research Foundation of Korea under
Grant No.\ NRF-2013R1A1A2A10007294.


\end{document}